\documentclass{epl}
\title{Fluorescence from doubly driven four-level atoms - A density
Matrix approach.}
\author{Andal Narayanan, R.Srinivasan, Ashok Vudayagiri, 
Uday Kumar Khan and Hema Ramachandran}
\institute{Raman Research Institute, Sadashivnagar, Bangalore, INDIA 560 080}
\begin{document}
\maketitle
\begin{abstract}
The unusually narrow features in the  fluorescence from  
 $^{85}Rb$ driven by the cooling and repumper laser fields,
reported in \cite{exp}, 
are explained on the basis of a four-level density matrix calculation. 
Quantum effects alter the efficiency of atom 
transfer by the probe (repumper)
laser to the levels connected by the pump (cooling) laser. 
This, combined with the double resonance condition \cite{exp},
leads to velocity selection from co-propagating and counter propagating
pump and probe beams resulting in narrow fluorescence peaks from 
a thermal gas at room temperature.
\end{abstract}

\section{Introduction}

	Multilevel atoms, under the action of multiple fields, display a variety of phenomena 
brought about by the ac Stark splitting of electronic levels and by interference due to
induced coherences. These manifest themselves as splitting of peaks, inhibition of absorption
or emission, narrow windows of transparency, etc. 
Theoretical work on this front have considered $\Lambda$, V and ladder type of
three level systems \cite{theor}. The presence of additional levels increases multifold the possibilities of
interference phenomena that exist. The nested $\Lambda$ and the N systems have been examined to 
some extent \cite{nsys}. 
	In this paper we examine an "inverted N" system using the density matrix formalism. Recent 
experimental observations in $^{85}Rb$ \cite{exp} have prompted the 
study of this four-level system under the
action of two fields.  

\section{Experiment}
\indent The experiment in \cite{exp} consisted of driving a collection of $^{85}Rb$ atoms with
two fields- a strong "cooling" field and a weak "repumper" field. 
One of the lasers, the cooling (pump) laser was held at a fixed detuning 
$\delta_c$ from the conventional cooling transition 
\footnote{ The cooling and repumper beams are used in the sense as defined
in ~\cite{exp}.}
($5S_{1/2}, F = 3 \rightarrow 5P_{3/2}, F'=4'$) and the repumper (probe)
laser was scanned with its detuning ($\delta_r$) from  $5S_{1/2}, F = 2 \rightarrow 5P_{3/2}, F'=3'$ level varying over the entire manifold, 
F=2 $\rightarrow$ F'=1',2',3' (See Fig. 1)
\footnote{Hyperfine levels of $5S_{1/2}$ are denoted unprimed
while that of $5P_{3/2}$ are denoted primed. }.
Narrow fluorescence peaks were observed at definite 
$\delta_r$ for a given $\delta_c$.  
While one would expect a  broad fluorescence, 
with the width indicative of the temperature
of the gas, the experiment 
showed fluorescence peaks just 30MHz in 
width, much smaller than their Doppler width of 500MHz. 
This was explained on the basis of 
a double resonance model which maximised fluorescence whenever
the atom found both the cooling and repumping lasers  
on resonance with their respective transitions. A simple analytical 
treatment based on the above model showed that atom transfer efficiency
by the repumper laser to the levels connected by the cooling laser 
was maximised under the double resonance condition and gave rise 
to narrow fluorescence peaks.\\
\indent In this paper, we present a rigorous treatment of the 
same problem using a four level density matrix formalism.
Our treatment differs from the earlier studies of four-level density
matrix \cite{earlier} in that 
(a) it considers an inverted N system
and (b) it takes into account the motion of the atoms in co- and counter-propagating 
geometries with respect to the cooling and repumper beams.
\begin{figure}
\onefigure{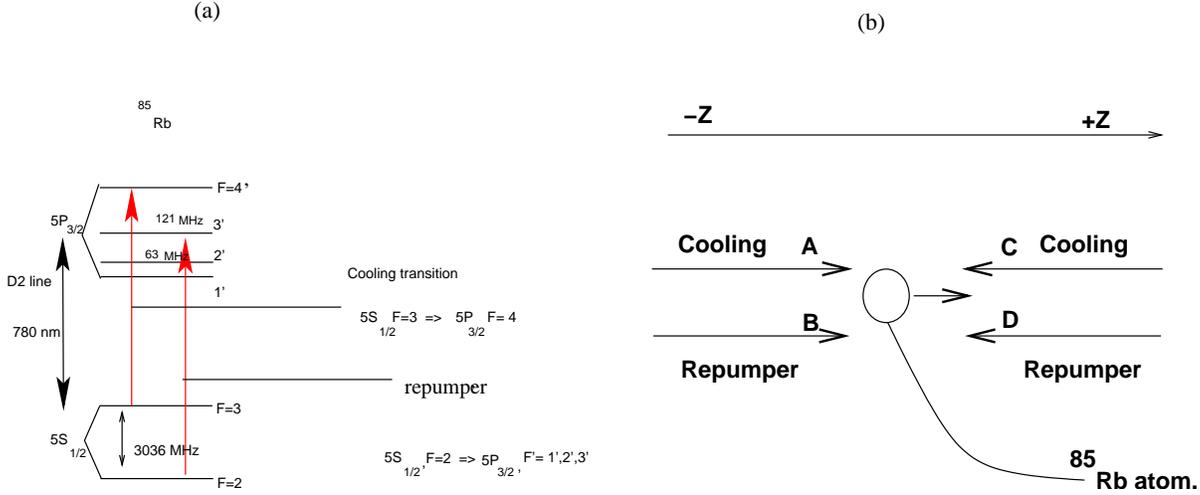}
\caption{(a): Energy level diagram of $^{85}Rb$. (b):A one dimensional configuration (along {\bf Z})
of the cooling and repumper beams with the
$^{85}Rb$ atom taken move along $+z$ direction
}
\end{figure}
\section{Four Level Density Matrix}
The four levels under consideration are: the two ground hyperfine levels
F=2,3 and the two excited levels F'=3',4'.
 For simplicity, we have considered an one dimensional situation 
where the two driving fields are in the $\pm z$ direction and the atom is
moving along the z direction with a velocity $\vec{v}$. 
The 
detuning of the cooling laser taking the Doppler effect into account 
is $\Delta_c$ = $\delta_c - \vec{k_c}.\vec{v}$  and
the repumper laser's detuning is 
$\Delta_r$ = $\delta_r - \vec{k_r}.\vec{v}$ where, $\delta_c$ and $\delta_r$
are the detunings of the laser in the laboratory frame.
The total Hamiltonian for the system consisting of the atom and the
light fields is written in the interaction picture as 
\begin{equation}
H = H_0 + H_I
\end{equation} 
where $H_0$ is the Hamiltonian for the bare atom and $H_I$ is the 
atom-light interaction Hamiltonian. They are given as
\begin{eqnarray*}
H_0 = \hbar \omega_2 |2><2| + \hbar \omega_3 |3><3| + \hbar \omega_{3'} |3'><3'| +
      \hbar \omega_{4'} |4'><4'|
\end{eqnarray*}
and
\begin{eqnarray*}
H_I =  -\frac{\hbar}{2}[ \Omega_{34'} |3><4'| \exp{(-i\omega_{LC} t)} + 
\Omega_{33'} |3><3'| exp{(-i\omega_{LC} t)} + 
\Omega_{23'} \exp{(-i\omega_{LR} t)} |2><3'| + 
H.C ]
\end{eqnarray*}
Here the $\hbar \omega_i$ represent the energies of 
the levels as represented in Figure 1, with $\hbar \omega_{2}$ taken
as zero, $\omega_{LC}$ and $\omega_{LR}$ the
frequencies of the cooling and repumper laser beams and 
$\Omega_{ij'}$ is the Rabi frequency connecting the levels $i$ and $j'$.
The total Hamiltonian can be written in matrix form as follows
\[ H = \left |\begin{array}{llcl}
\hbar \omega_2 & 0 & -\frac{\hbar}{2}\Omega_{23'} e^{(-i\omega_{LR} t)} & 0 \\
0 & \hbar \omega_3 &-\frac{\hbar}{2} \Omega_{33'} e^{(-i\omega_{LC} t)} & 
 -\frac{\hbar}{2} \Omega_{34'} e^{(-i\omega_{LC} t)}  \\
-\frac{\hbar}{2} \Omega_{23'}  e^{(i\omega_{LR} t)} & 
-\frac{\hbar}{2} \Omega_{33'}  e^{(i\omega_{LC} t)}
 & \hbar \omega_{3'} & 0 \\
0 & -\frac{\hbar}{2} \Omega_{34'} e^{(i\omega_{LC} t)} & 0 & \hbar \omega_{4'} 
\end{array} \right| \]
where the rows and columns correspond to levels $2,3,3'$, 4' in sequence.
The dynamics of the system described by this Hamiltonian can be studied using 
the density matrix $\rho =\sum \rho_{ij} |i><j|$.
The time evolution of the density matrix $\rho$ is given
 by the Liouville equation
\begin{equation}
\frac{d\rho}{dt} = -\frac{i}{\hbar} [H,\rho] -\frac{1}{2} \{\Gamma, \rho \}  \label{lio}
\end{equation}
with 
\begin{equation}
\Gamma_{ij} = 2 \gamma_{i' \rightarrow j} \delta_{ij'}
\end{equation}
where $\gamma_{i' \rightarrow j}$ being the spontaneous decay 
rate from the $j^{th}$ level to the $i'^{th}$ level. 
\footnote{where $\Gamma_{ij'}$ is the decay rate of the 
level j to the level i'. This is however a diagonal matrix}
The rate equations of the four levels for an atom moving with a
velocity $v$ 
are derived under the rotating wave approximation. 
They are
\begin{eqnarray}
\frac{d\rho_{11}}{dt} = -\frac{i}{2}[ \Omega_{23'} \rho_{31} - \Omega_{23'}^{*} \rho_{13}] + 2\gamma_{3'2} \rho_{33'} \\
\frac{d\rho_{12}}{dt}  = -i(\Delta_{c3'} - \Delta_{r}) -\frac{i}{2}[ \Omega_{23'} \rho_{32} - \Omega_{33'}^{*} \rho_{13} - \Omega_{34'}^{*} \rho_{14} ]\\
\frac{d\rho_{13}}{dt} = (i(\Delta_r)- (\gamma_{3'2} + \gamma_{3'3}) )\rho_{13}
-\frac{i}{2}[ \Omega_{23'}\rho_{33} 
- \Omega_{23'}^{*} \rho_{11} -\Omega_{33}^{*} \rho_{12}]\\
\frac{d\rho_{14}}{dt} = -i[(\Delta_{c3'} - \Delta_c -\Delta_r) -  \gamma_{4'3}]\rho_{14} - \frac{i}{2}[\Omega_{23'} \rho{34'} -\Omega_{34'}\rho_{12}]\\
\frac{d\rho_{21}}{dt} = i(\Delta_{c3'} - \Delta_r)\rho_{21}-\frac{i}{2}[\Omega_{33'} \rho_{31} +\Omega_{34'} \rho_{41} -\Omega_{23'}\rho_{23}]\\
\frac{d\rho_{22}}{dt} = -\frac{i}{2}[\Omega_{33'}\rho_{32} + \Omega_{34'} \rho_{42} -\Omega_{33'}\rho_{23} - \Omega_{34'} \rho_{24}]+2\gamma_{3'3}\rho_{33}
+2\gamma_{4'3} \rho_{44})\\
\frac{d\rho_{23}}{dt} = (i\Delta_{c3'} - (\gamma_{3'2} +\gamma_{3'3}))\rho_{23} -
\frac{i}{2}[\Omega_{33'}\rho_{33} + \Omega_{34'}\rho_{43} - 
\Omega_{23'}\rho_{21} -\Omega_{33'}\rho_{22}]\\
\frac{d\rho_{24}}{dt} = (i\Delta_c - \gamma_{4'3})\rho_{24} -\frac{i}{2}[\Omega_{33'} \rho_{34} +\Omega_{34'}\rho_{44} -\Omega_{34'}\rho_{22}]\\
\frac{d\rho_{31}}{dt} = (-i\Delta_r - (\gamma_{3'2} +\gamma_{3'3}))\rho_{31} -\frac{i}{2}[\Omega_{23'}^{*} \rho_{11} + \Omega_{33'}^{*}\rho_{21} 
- \Omega_{23'}^{*} \rho_{33}]\\
\frac{d\rho_{32}}{dt} = (-i\Delta_{c3'} -(\gamma_{3'2} + \gamma_{3'3}))\rho_{32} 
-\frac{i}{2}[\Omega_{33'}^{*}\rho_{22} -\Omega_{33}^{*}\rho_{33} 
- \Omega_{34'}\rho_{34}]\\
\frac{d\rho_{33}}{dt} = -2(\gamma_{3'2} +\gamma_{3'3})\rho_{33} 
-\frac{i}{2}[\Omega_{23'}\rho_{13} +\Omega_{33'}^{*} \rho_{23} 
-\Omega_{23'}\rho_{31} ] \\
\frac{d\rho_{34}}{dt} = [(i\Delta_{c3'}-\Delta_c)-(\gamma_{3'2} +\gamma_{3'3} + \gamma_{4'3})]\rho_{34} -\frac{i}{2}[\Omega_{23'}^{*} \rho_{14} +\Omega_{33'}^{*}\rho_{24} -
\Omega_{34'} \rho_{32}]\\
\frac{d\rho_{41}}{dt} = [(i\Delta_{c3'}-\Delta_c -\Delta_r)-\gamma_{4'3} ]\rho_{41}-
\frac{i}{2}[\Omega_{34}^{*} \rho_{21} -\Omega_{23'}^{*} \rho_{43}]\\
\frac{d\rho_{42}}{dt} = [(i\Delta_c-\gamma_{4'3}) ]\rho_{42}-
\frac{i}{2}[\Omega_{34}^{*} \rho_{22} -\Omega_{33'}^{*} \rho_{43}-\Omega_{34'}^{*} \rho_{44}]\\
\frac{d\rho_{43}}{dt} = [-i(\Delta_{c3'}-\Delta_c)-(\gamma_{3'2}+\gamma_{3'3}+\gamma_{4'3}) ]\rho_{43}-
\frac{i}{2}[\Omega_{34'} \rho_{23} -\Omega_{33'} \rho_{42}-\Omega_{34'}^{*} \rho_{23'} \rho_{41}]\\
\frac{d\rho_{44}}{dt} = -\frac{i}{2}[\Omega_{34'}^{*} \rho_{24} - \Omega_{34'}
\rho_{42} - 2\gamma_{4'3}\rho_{44}] 
\end{eqnarray}
For each velocity $\vec{v}$ steady state values of $\rho_{ij}$
 are obtained by numerically  solving 
the above
rate equations for various values of $\delta_c$ and $\delta_r$,
subject to the constraint 
$\sum_{i=1}^{4} \rho_{ii} = 1$.
Thus for an atom with a velocity $\vec{v}$ we obtain the population of 
each level and coherence between various levels for different values
of $\delta_{r}$ and $\delta_c$.
The fluorescence emitted by atoms with a velocity $\vec{v}$ is given by
\begin{equation}
Fluorescence(\Delta_c,\Delta_r) = \Gamma_{4'3} \rho_{4'4'} + \Gamma_{3'3} \rho_{33'} + \Gamma_{3'2} \rho_{3'3'}
\end{equation}
As the detector collects fluorescence from atoms which are in thermal 
motion, we take the average of 
each $\rho_{ij}$ value over the range of velocities,
weighted by the one dimensional Maxwellian velocity distribution.
The fluorescence 
calculated as above as a function of $\delta_r$ 
and the corresponding
experimental data are given in Fig. 2 for a cooling laser
detuning $\delta_c = -162$ MHz. 
\begin{figure}
\onefigure{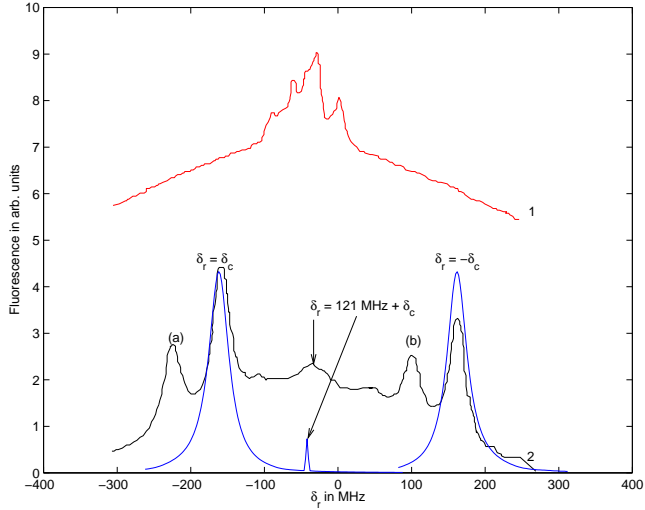}
\caption{ Trace 1 shows the saturation absorption spectrum of the repumper.
Trace 2 shows the experimental curve (black) and the corresponding theoretical curve (blue).
See text for explanation on the (a) and (b) peaks in the figure.
}
\end{figure}
A general agreement between the results of our calculation
and that of the experiment is seen. \footnote {Peaks (a) and (b) seen in the experiment
arise due to the transitions from $F'=2'$ level which has not been included in our density
matrix}
The individual features will be discussed in detail below.
\section{Discussion}
Consider a situation shown in Fig. 1b, when the cooling and repumping
beams are both along $\pm z$ directions. Initially let us consider the atoms 
to be at rest. 
For a given detuning $\delta_c$ of the cooling (pump) laser, 
we should get fluorescence peaks corresponding to the Autler-Townes (AT)
dressed states of $F'=3'$ at the probe (repumper) detunings 
\cite{Agar}
\begin{equation}
\delta_{r\pm} = \frac{\delta_{c3'}}{2} \pm \sqrt{(\delta_{c3'}^2  + \Omega^2)}/2 
\end{equation}
Here $\delta_{c3'}$ = $2\pi*121$MHz + $\delta_c$ 
denotes the detuning of the cooling laser from
$3 \rightarrow 3'$ transition, $121$ MHz being 
the level spacing between $F'=3'$ and $4'$ levels and $\Omega$ its Rabi frequency.
For the specific case shown in figure 2, for a 
cooling laser detuning $\delta_c = -162 MHz$ and for small $\Omega$, 
the Autler-Townes peak positions for these zero velocity atoms are 
$\delta_{r\pm} \approx 0, -41$ MHz. 
\begin{figure}
\onefigure{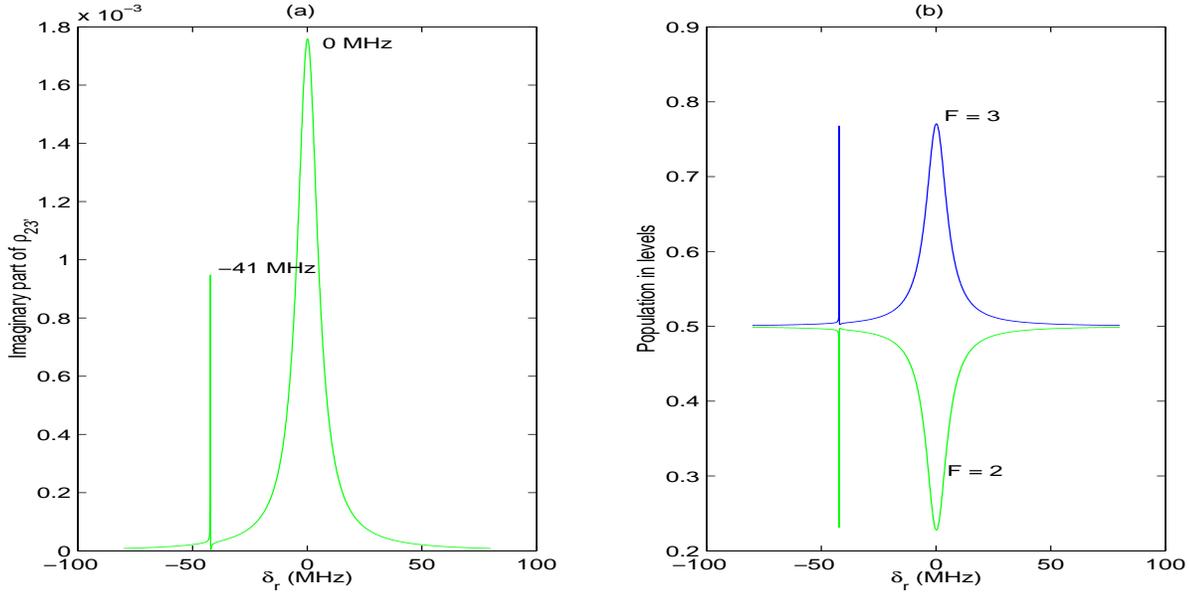}
\caption{The imaginary part of $\rho_{23'}$ vs the positions of the probe absorption $\delta_r$
showing that absorption of the repumper (probe)
light due to Autler -Townes splitting of the
level $F'=3'$ occurs at the detunings $0 MHz$ and $-41 MHz$.
(b) $\rho_{33}$ (blue) and $\rho_{22}$ (green) as functions of $\delta_r$.
}
\end{figure}
This is confirmed from our density matrix calculations.
At these detunings of the probe 
its absorption is maximised and repumping most efficient
as seen from Figs. 3(a) and 3(b) which give $\rho_{23'}$, $\rho_{33}$ and $\rho_{22}$
as functions of $\delta_r$.\\
\indent  Consider now an atom in motion with a velocity $+v$,
along the $+z$ direction. If
$\delta_c < 0$  this atom predominantly absorbs 
from the cooling beam coming towards it  
(C).
Since
the repumper laser is scanned, depending on the sign of $\delta_r$ the atom
absorbs either from B or from D. So for a given $\delta_c$,
altogether we get four peaks, 
a Autler-Townes pair each shifted due to the Doppler effect,
 depending upon whether absorption takes place
from the repumper beam B or from repumper beam D.  
The same holds for $\delta_c >0$ for atoms with
velocity $-v$ .\\
\indent The AT doublet positions as given in equation (21) when applied to
atoms in motion and,
for small $\Omega$ and $\delta_c < 0$ are
\begin{eqnarray}
\delta_{r-} - (\pm kv) \approx 0  \label{at1} ; &
\delta_{r+} - (\pm kv) \approx \delta_{c3'} - (\pm kv)  \label{at2}
\end{eqnarray}
As the repumper is scanned, maximum transfer of population
from F =2 to F= 3 will occur when these conditions are satisfied. 
Now, as mentioned in \cite{exp} and as will be shown shortly,
the fluorescence is predominantly from $4'-> 3$.  However,
not all atoms in F =3 will be in resonance with the cooling laser; only
a small 
velocity class around $v_c = \delta_c/k$
will give rise to fluorescence. 
This is the mechanism that gives rise to 
narrow velocity selection from a hot gas.
This velocity selection effect is confirmed by our calculations and is
shown in Fig. 4 which gives $\rho_{4'4'}$ and $\rho_{3'3'}$ as functions
of the velocity of the atom for $\delta_c = \delta_r$= -162 MHz. 
\begin{figure}
\onefigure{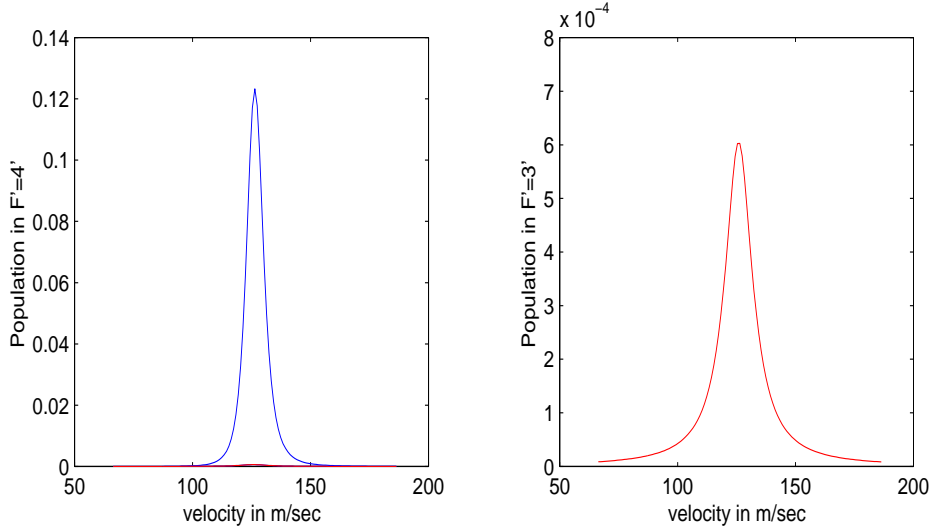}
\caption{Populations in the upper levels $F'=3$ and $F'=4'$. The very low 
population of $F'=3'$ in (a) is shown magnified in (b).
}
\end{figure}
We see that the population in F'=4' ($\rho_{4'4'}$) is two orders 
more than that in 
F'=3' ($\rho_{3'3'}$) and
it peaks at the critical velocity $v_c = 126 m/sec$. Therefore only
for velocities around $v_c$ will the fluorescence from $F'=4'$ be maximised,
showing that indeed the double resonance condition selects a narrow velocity
class for fluorescence.\\
\indent The Autler-Townes peak positions, as given by  equations 
($~\ref{at2}$), with the constraint that double resonance 
should also be satisfied for maximum fluorescence, require that
\begin{eqnarray}
\delta_{r-}  = \delta_c     \label{pos1} ; &
\delta_{r+} = 121 MHz + \delta_c  \label{pos2} 
\end{eqnarray}
when repumper absorption takes place from D  
and
\begin{eqnarray}
\delta_{r-} = -\delta_c  \label{pos3} ; &
\delta_{r+} = 121 MHz - \delta_c (v = v_c); & \delta_{r+} = 121 MHz + \delta_c +2kv (v \neq v_c)
 \label{pos4}
\end{eqnarray}
when repumper absorption takes place from B. \\
The same conditions result for atoms  
with a  velocity $-v$, for $\delta_c > 0$.
It has been estimated in \cite{Agar} that 
the peak at $\delta_{r-}$ is broad and the one at $\delta_{r+}$ is narrow. 
The width of the AT peaks decide the prominence of a fluorescence peak as it 
decides the number of atoms participating in the fluorescence.
Thus for peak positions $\delta_{r-}$ = $\delta_c$ \& $-\delta_c$ we expect a large fluorescence 
whereas for the peak at $\delta_{r+} = 121 MHz + \delta_c$ we expect a much smaller fluorescence
as is indeed seen from experiment and from our calculations (Fig. 2).
As mentioned in \cite{exp} this peak is well resolved only at large
detunings.\\
\indent  The peak around $\delta_{r+}$ =
$121$ MHz + $\delta_c + 2kv$ is absent both in experiment
and in our density matrix calculation.  For this case,
the repumper absorption takes place from B whereas the cooling is absorbed from C.
When the absorption takes place from counter-propagating cooling and repumper beams
the velocity class satisfying the double resonance is severely restricted.
In fact only for $v=v_c$ will the double resonance condition be satisfied. 
Atoms with 
$v \neq v_c$ will see the cooling and repumper to be shifted by different detunings 
and hence will not contribute to the fluorescence. 
Thus the peaks
resulting from this configuration will not be resolvable as only a very small
number of atoms contribute to it.
The severely restricted velocity range at resonance 
resulting in the absence of the peak at  
$121 MHz + \delta_c + 2kv$  
is illustrated in Fig. 5b which is obtained from the density matrix calculation. 
As we see here, only a small number in a very narrow velocity range contribute to
population in F'=4' in contrast to the case $\delta_{r-} = -\delta_c$ (Fig. 5a).\\
\begin{figure}
\onefigure{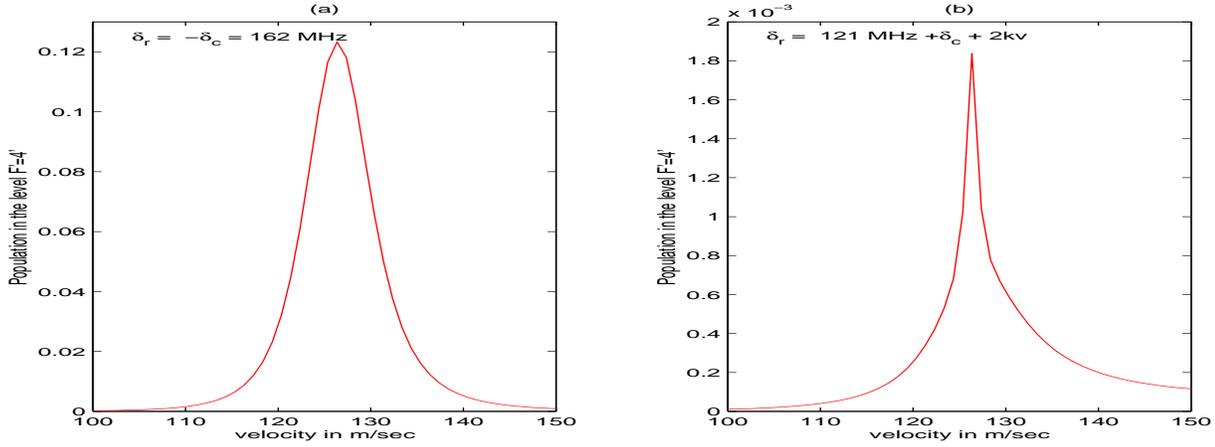}
\caption{ Population in the level $F'=4'$ as a function of the velocity of atoms
for the Autler-Townes peaks at
(a) $\delta_r = -\delta_c = 162 MHz$ (b) $\delta_r = 121 MHz + \delta_c +2kv$.
}
\end{figure}
\indent The peaks marked (a) and (b) in Figure 2 are the peaks corresponding
to the AT levels of $F'=2'$ at $\delta_{r-} = \delta_c$ \& $-\delta_c$ corresponding to
absorption from counter and co-propagating repumper beams.
Our calculation has not reproduced this peak, as the level F'=2' was
not included in the density matrix. Nevertheless the peak positions can be
simply calculated from the corresponding peaks for F'=3' by shifting
them by $-63 MHz$. The peak
at $\delta_c + 121 MHz$ falls at the same $\delta_{r+}$ for both $F' = 2'$ and 
$F'=3'$.\\
\indent The theoretically calculated widths of the fluorescence 
peaks match the experimentally observed narrow width of about $30 MHz$. 
It should be emphasised that this does not arise due to the 
cooling of atoms in the optical molasses like configuration but due
to the velocity selection as discussed above. 
These
fluorescence peaks are experimentally seen to be narrow even for
blue detunings of the cooling beam where no cooling occurs and the atoms are at
{\it room temperature}.
The corresponding Doppler width
is several hundred MHz at this temperature.
The additonal effects of cooling on these fluorescent
peaks in a Doppler free beam geometry will be discussed elsewhere. 
\section{Conclusions}
The experimental observations of \cite{exp} has been explained 
using a four-level
density matrix formalism. The theory finds that the fluorescence 
peaks are given rise by the Autler-Townes (AT) doublets of $F'=3'$
and $F'=2'$ for atoms around 
a particular velocity class. The theory gives the three AT peaks of $F'=3'$
at 
positions borne out by the experiment. The theory also explains the absence
of the fourth peak.
The widths and heights 
of the peaks agree quite well with the experimentally obtained
widths and heights.

\end{document}